\documentclass[aps,pre
,twocolumn,groupedaddress,showpacs,nofootinbib]{revtex4-1}
\begin{document}
\preprint{CERN-PH-TH/2012-348}

\title{Structure of Amplitude Correlations in open chaotic Systems}


\author{Torleif E. O.   Ericson}
\email{Torleif.Ericson@cern.ch}
\affiliation{Theory Group, Physics Department, CERN, CH-1211 Geneva, Switzerland}



\begin{abstract}
The Verbaarschot-Weidenm\"uller-Zirnbauer model 
is believed to correctly represent  the correlations  of two S-matrix elements  for an open quantum chaotic system, but the solution has considerable complexity and is presently only accessed numerically.  Here a procedure is developed to deduce its features  over the full range of the parameter space in a transparent and simple analytical form preserving accuracy to a considerable degree.The bulk of the VWZ correlations are described by the Gorin-Seligman  expression for the 2-amplitude correlations  of the Ericson-Gorin-Seligman  model.  The  structure of the remaining correction factors for correlation functions is discussed with special emphasis of the r\^ole of the level correlation hole both for inelastic and elastic correlations.
\end{abstract}

\pacs{05.45.Mt,24.60.Dr,24.60.-k}
\maketitle

\section{Introduction \label{Intro}}
 The Verbaarschot-Weidenm\" uller-Zirnbauer (VWZ) model   for chaotic scattering for an open system \cite{VER85} is generally believed to be a realistic example of such processes\cite{PLU12,DIE10A}.  It is based on a Random Matrix Model for which the random Hamiltonian is dynamically coupled to the external channels.  For the correlations of two S-matrix elements it gives an explicit general solution, which unfortunately is very complex.  In spite of the importance of this result, the dominant physical features  are mostly not  transparent.  Hence they cannot  be reliably exploited in simplified form for application to higher order correlation functions such as cross-section correlations for which the corresponding VWZ solutions are not known presently.
 The VWZ   amplitude correlations can presently be evaluated only using numerical methods with specified values for the numerous parameters \cite{GOR02,DIE11}, but for special cases \cite {MUE90}. 
Approximate expressions exist  for small correlation times, but the  convergence radius of the expansion is small \cite{WEI84,VER86}. \\
The properties of higher order correlation functions in the VWZ model are largely unaccessible with present approaches and their properties are unexplored with few exceptions. The cross-section correlations (4-point  amplitude correlations) have been obtained for a vanishing correlation time \cite{DAV88,DAV89,DIE10}. Their behavior is however only accessed numerically. 
A realistic study of these correlation functions requires then a simplified approach in which the properties of the amplitude correlations closely mirror those of the VWZ model. In the following we will demonstrate that the model due to  Ericson-Gorin-Seligman (EGS) \cite{ERI60,GOR02} gives a good general representation of  the VWZ solution for amplitude correlations, but with some characteristic deviations.
Its explicit solution is in some respects more general than the one of the VWZ model and has been used in a study to be published to explore the structure of cross-section correlations  both for near-chaotic and for chaotic scattering from open systems\cite{DIE12}.  \\
The EGS model is inspired by reaction theory such as in Ref.  \cite{FES62}.
 It approximates the S-matrix   by a sum of complex poles, the positions of which are assumed not to be influenced by the  coupling to the external channels. In its main version   these poles have partial width amplitudes with a normal (Gaussian) distribution; the pole positions are assumed to have the same spacing distribution as for the  Random Matrix Model \cite {MEH67}, i. e., generated by matrix elements with a normal distribution. Although the EGS model implements a number of features of the VWZ one, it differs  in particular by not accounting fully for the dynamical feed-back of the open channels on the spacing distribution\cite{GOR97}. In addition, the  price for the solvability of the EGS model is that it violates unitarity sum-rules, which potentially may produce unrealistic properties. It is shown however in Ref. \cite{DIE12}  that  these restrictions are of unimportant in most situations;  this conclusion is concretely vindicated by the results of the present article.    As a check on the effectiveness of this approach the Gorin-Seligman (GS) results for the 2-point correlations of the EGS model\cite{GOR02}  have been  numerically compared to the ones in the VWZ model\cite{DIE11}. The finding is that this model with normally distributed partial width amplitudes describes the gross features of the VWZ model surprisingly well for a variety of conditions. This suggests that the structure of the GS result might be used as a guide to develop an accurate and transparent analytical  approximation to the VWZ results.  This is the goal of the present article. It  gives a valuable insight in addition to the accurate numerical results  for the 2-point function. This  permits one to  identify  explicitly the origin and nature of the differences  of the VWZ result and the GS one.  The understanding of this point gives confidence that the EGS model gives a reliable approximate description of the chaotic cross-section correlations \cite{DIE12}. 
\\
The strategy, therefore, will be to extract the main variation of the key VWZ result with the  GS solution as guide to identify appropriate expansion parameters for the deviations. Using this information a natural procedure   is developed for obtaining  explicit closed expressions for the additional factors over the entire parameter space.\\
\indent
In Sec. \ref{SEC:II}  the standard VWZ 2-point function (8.10) in Ref. [1]
  is converted to the time representation and expressed more conveniently in terms of  the time variable $\tau $.   This representation is technically easier to handle than the energy one, since it gives products and not folding integrals.  In addition, it is  immediately apparent that there are no singularities in the integrals. The basic correlation functions and their notation are defined here.\\
  \indent
  In Sec. \ref{SEC:III} some basic identities associated with VWZ integral are given as well as their relations to the level spacing correlations. This section also defines notations for  the weighted functions generated by the integrations.  \\
 \indent
 Sec. \ref{SEC:IV} describes the construction of the small expansion parameters, the basic expansion and the  expansion of  the factors  coupling  to the observed channels referred to as the 'leads'.  It gives in addition the structure of the general expansion term.\\
 \indent
 Sec. \ref{SEC:V} gives the explicit expressions for small and large $\tau $ as well as a convenient interpolation connecting the two regions.\\
 \indent
 Sec. \ref{SEC:Disc} is devoted to a discussion of the characteristic features of the amplitude correlation functions in the light   of the results. It also discusses possible generalizations and some practical consequences.  

\begin{widetext}
\section{The VWZ Correlation Function \label{SEC:II}}
  The key result  in Ref. [1] is
   Eq. (8.10), which gives the  correlation function for two conjugate S-matrix amplitudes $S(E(1))$ and $S^*(E(2)) $ at energies $E(1,2)$ differing by $\epsilon =E(1)-E(2)$. The system is assumed to have no secular energy dependence with $[E(1)+E(2)]/2$ and threshold effects in the different channels are neglected. The observed channels (the 'leads') are denoted by $a,b,c,d$ while open channels without restrictions are denoted by $e$ with corresponding transmission coefficients $T_e$. Ensemble averages are denoted by $<~>$. The energy scale has units  such that the average level  spacing  $d=1$.

The general VWZ correlation function in the energy representation is given by the triple integral
  \begin{eqnarray}
 \hspace{-0.3in}&&  
C_{ab;cd}(\epsilon )
\equiv  <S_{ab}(E(1))S^* _{cd}(E(2))> - <S_{ab}><S ^* _{cd}>=  \label{Grassman}	 \\  \nonumber
   \hspace{-0.3in}&&  
  {1\over 8} \int ^{\infty }_0 d\lambda _1\int ^{\infty }_0 d\lambda _2 \int ^{1 }_0 d\lambda  {(1- \lambda )\lambda |\lambda _1-\lambda _2| \over 
  \left[(1+\lambda _1)\lambda _1(1+\lambda _2)\lambda _2\right]^{1/2}(\lambda +\lambda _1)^2(\lambda +\lambda _2)^2}\times \\ \nonumber
   \hspace{-0.3in}&& 
 exp \left[-( i \pi \epsilon ) (\lambda _1 +\lambda _2 + 2\lambda )\right] \prod _e {(1-T_e\lambda ) \over (1+T_e\lambda _1)^{1/2}(1+T_e\lambda  _2)^{1/2}} \times \\  \nonumber
     \hspace{-0.3in}&& 
    \Big\{\delta _{ab}\delta _{cd} <S_{aa}><S ^*_{cc}> T_aT_c\left({\lambda _1\over 1+T_a\lambda _1}+
      {\lambda _2\over 1+T_a\lambda _2}+{2\lambda \over 1-T_a\lambda }\right)
      \Big({\lambda _1\over 1+T_c\lambda _1}+{\lambda _2\over 1+T_c\lambda _2}+
     {2\lambda \over 1-T_c\lambda }\Big)+
      \\  \nonumber  \hspace{-0.3in}&& 
      \left(\delta _{ac}\delta _{bd}+ \delta _{ad}\delta _{bc}\right)T_aT_b       \nonumber  
      \Big[  {\lambda _1(1+\lambda _1)\over (1+T_a\lambda _1)(1+T_b\lambda _1)}+{\lambda _2(1+\lambda _2)\over (1+T_a\lambda _2)(1+T_b\lambda _2)}+{2\lambda (1-\lambda )\over (1-T_a\lambda) (1-T_b\lambda) }  \Big] \Big\}
   \end{eqnarray}   
   The normalization is determined by the triple integral of a weight factor and gives no singular contribution from the region near the origin
  \begin{eqnarray}
 \hspace{-0.3in}&&  
  {1\over 8} \int ^{\infty }_0 d\lambda _1\int ^{\infty }_0 d\lambda _2 \int ^{1 }_0 d\lambda  {(1- \lambda )\lambda |\lambda _1-\lambda _2| \over 
  \left[(1+\lambda _1)\lambda _1(1+\lambda _2)\lambda _2\right]^{1/2}(\lambda +\lambda _1)^2(\lambda +\lambda _2)^2}=1~~~~~\label{Eq:norm}
  \end{eqnarray}

  While these expressions present nicely in these variables, it is not the most convenient representation for applications.  
The equivalent correlation function in the time representation is obtained by taking the Fourier transform using that  $\int _{-\infty }^{\infty} d\epsilon \exp[2\pi i x\epsilon  ] =\delta (x)$, where the energy is measured in units of the average level spacing $d$.
 The time $\tau \geq 0$ is measured in standard units $2\pi /d$  as , e. g.,  in Ref. [12], Eq. (8).  This means in particular that the condition for overlapping levels $\Gamma >d$ corresponds to $\sum _eT_e>2\pi  $ .
  It is convenient to make the variable substitutions $\lambda _{1,2} =2\tau x_{1,2}$ and $\lambda =\tau x$ .  
  These transformations give immediately the Fourier transform as
      \begin{eqnarray}
 \hspace{-0.3in}&&  
\tilde C_{ab;cd}(\tau)
\label{Eq:time}\equiv  \int _{-\infty }^{\infty} d\epsilon  \exp (2\pi i\epsilon \tau ) C_{ab;cd}(\epsilon)=
    \\   \hspace{-0.3in}&&  
 \int ^{\infty }_0 dx _1\int ^{\infty }_0 dx _2 \int ^{min(1,1/\tau ) }_0 dx  {(1-\tau  x )\over  (1+2\tau x _1)^{1/2}(1+2\tau x _2)^{1/2}} {x |x _1-x _2| \over  
(x _1x _2)^{1/2}( x +2x _1)^2(x +2x _2)^2}\nonumber   \times  \\ 
      \nonumber 
   \hspace{-0.3in}&& 
\delta  \left[1- (x _1 +x _2 + x )\right] \prod _e {(1-\tau T_ex) \over (1+2\tau x _1T_e)^{1/2}(1+2\tau x_2T_e)^{1/2}}  \times  \\ 
      \nonumber 
     \hspace{-0.3in}&& 
    \Big\{\delta _{ab}\delta _{cd} <S_{aa}><S ^*_{cc}> 2\tau
     T_aT_c\left({x _1\over 1+2\tau T_a x_1}+{ x _2\over 1+2\tau T_a x _2}+{x \over 1-\tau T_a x }\right)         \left({x _1\over 1+2\tau T_c x_1}+{ x _2\over 1+2\tau T_cx _2}+{x \over 1-\tau T_cx }\right) +
     \\   \nonumber        \hspace{-0.3in}&&
      \left(\delta _{ac}\delta _{bd}+ \delta _{ad}\delta _{bc}\right)T_aT_b      \Big[  {x _1(1+2\tau x_1)\over (1+2\tau T_a x _1)(1+2\tau T_b x _1)}+{\ x _2(1+2\tau x _2)\over (1+2\tau T_a x _2)(1+2\tau T_b x _2)}+{x (1-\tau x )\over (1-\tau T_ax) (1-\tau T_bx) }  \Big] \Big\}
  \end{eqnarray}
  The expression (\ref{Eq:time}) is the basis for the following. 
   It is convenient to work with the correlation functions $\tilde {C}_{aa;cc}^{1} (\tau  ) $ and $\tilde {C}_{ab;ab}^{2} (\tau  ) $ defined from this expression: 
      \begin{eqnarray}
     \hspace{-0.3in}&&  
     \tilde C_{ab;cd}(\tau)
 \equiv    \delta _{ab}\delta _{cd} \tilde {C}_{aa;cc}^1 (\tau  ) +  \left(\delta _{ac}\delta _{bd}+ \delta _{ad}\delta _{bc}\right)  \tilde {C}_{ab;ab}^2 (\tau  )~ \label{Eq:C1,2-1tilde}
  \end{eqnarray}
  There are 3 basic  situations: elastic autocorrelations , correlations of two different elastic amplitudes  and inelastic autocorrelations. Other correlations between two S-matrix amplitudes vanish for the VWZ 2-point function. The relevant cases are explicitly
    \begin{eqnarray}
     \hspace{-0.3in}&&  
     \tilde C_{aa;aa}(\tau)~~~~~~~~~~~~~~~~~=~~~~~\tilde C^1_{aa;aa}(\tau)+2\tilde C^2 _{aa;aa}(\tau) \label{Eq:C1,2-1tildeA}~~~~~~~{\rm elastic ~autocorrelations}\\
  \hspace{-0.3in}&&  
    \tilde  C_{aa;cc}(\tau)~~~~~~~~~~~~~~~~~=~~~~~~\tilde C^1_{aa;cc}(\tau)  ~~~~~~~~~~~~~~~~~~~~~~~~{\rm elastic ~correlations ~with}~a\neq c
  \\
  \hspace{-0.3in}&&  
  \tilde  C_{ab;ab}(\tau)=\tilde  C_{ab;ba}(\tau)~~= ~~~~~\tilde C^2_{ab;ab}(\tau)~~~~~~~~~~~~~~~~~~~~~~~~~{\rm inelastic ~autocorrelations~with} ~a\neq b
  \end{eqnarray}
\section{Basic Identities \label{SEC:III}}
The following identities are crucial in the expansions for small and moderate $\tau$. They result 
in the limit of vanishing transmission coefficients and are central in deriving non-trivial approximations in the following.
Exact integrals are obtained from Eq. (\ref{Eq:time}), where the index $\tau $ on a quantity means that it is an averaged integral weighted by the parts of the integrands that are independent of the transmission coefficients as in Eq. (\ref{Eq:identitiesA}) below. In the limit  of vanishing transmission coefficients the correlation function $\tilde {C}_1(\tau  )$ defines the following function  \begin{eqnarray}
 \hspace{-0.3in}&&  
I(\tau  )\label{Eq:identitiesA}\equiv \\ 
      \nonumber    \hspace{-0.3in}&&  
 \int ^{\infty }_0 dx _1\int ^{\infty }_0 dx _2 \int ^{\, min (1,1/\tau ) }_0 dx  {(1-\tau  x )\over  (1+2\tau x _1)^{1/2}(1+2\tau x _2)^{1/2}} {x |x _1-x _2| \over  
(x _1x _2)^{1/2}( x +2x _1)^2(x +2x _2)^2}
\times\delta  \left[1- (x _1 +x _2 + x )\right]
\\  \hspace{-0.3in}&& \nonumber
 \end{eqnarray}
Here $I(\tau =0)=1$. One notes that the time-dependent factor formally appears as an additional channel $e$ with $T_e=1$.   We will in the following denote by $[f(x_1,x_2,x)]_{\tau }I(\tau )$ the integral with the same time-dependent integrand weighted by $f(x_1,x_2,x)$.  In Eq. (\ref{Eq:identitiesA}) this definition of $I(\tau )$ corresponds to the integration of the integrand multiplied by $1$ with $[1]_{\tau }\equiv 1$. 
 \begin{eqnarray}
 \hspace{-0.3in}&&   
     2\tau  I (\tau  )=1-b_1(\tau) ~~
     \label{Eq:I(tau)explicitA}
       \end{eqnarray}
      The last term in this expression for $I(\tau )$ is exactly  the Dyson level correlation function $b_1(\tau )$  \cite{DYS62} , while the first  term $1$ in Eq. (\ref{Eq:I(tau)explicitA}) comes from the self-correlation of a narrow level with itself.  Here  $b_1(\tau )=\int dsY_{2;1}(s)\exp (2\pi i\tau  s)$ of the corresponding level correlation in the energy representation. It has the explicit form~\cite{MEH67}
          \begin{eqnarray} 
 \hspace{-0.3in}&& 
{\rm For} ~0<\tau  <1~~  b_1(\tau )= 1-2\tau +\tau \ln [1+2\tau ] \simeq  ~1-2\tau +2\tau^2-2\tau ^3+8\tau ^4/ 3-4\tau ^5 ..
   \nonumber \\
 \hspace{-0.3in}&&  \label{Eq:Dyson}
 {\rm For} ~~~~~~~\tau  > 1~~   b_1(\tau )= -1+\tau  \ln \left[{2\tau +1 \over 2\tau -1}\right]~\simeq 0+(1/ 12)\tau ^{-2}+..~~\label{Eq:b-1}
   \end{eqnarray}
   \end{widetext}
   Here $b_1(1)\simeq 0.0986..$ and it vanishes rapidly for larger $\tau $. 
   This level correlation function is an explicit result in  the VWZ model found by Efetov and Verbaarschot \cite{EFE83,VER86}.

   A similar, but simpler, relation is obtained similarly from the correlation function $\tilde {C}_2(\tau  )$ in the limit of all $T_e=0$. In this case the origin is also a level correlation, but here only the self-correlation of the levels contributes and gives only the first term in Eq. (\ref{Eq:I(tau)explicitA}).
    \begin{eqnarray}
 \hspace{-0.3in}&&  
 [1+2\tau (x_1^2+x_2^2-(1/2)x^2)]_{\tau }I(\tau )=1 \label{Eq:x_1^2+x_2^2-(1/2)x^2A}
   \end{eqnarray}
  
   It is convenient to regroup the terms in Eq. (\ref{Eq:x_1^2+x_2^2-(1/2)x^2A}) using $x_1+     x _2+ x =1$ so as to display  the parameters $x_1x_2$ and $x$ which will serve as small expansion parameters in the following:
  \begin{widetext}
    \begin{eqnarray}
 \hspace{-0.3in}&&  
 x_1^2+     x _2^2-(1/2) x ^2\equiv 1  -2\left(x_1x_2+x-(1/4)x^2\right)
   \end{eqnarray} 
   With the expression (\ref{Eq:I(tau)explicitA}) for $I(\tau )$ in terms of the Dyson function 
   \begin{eqnarray}
 \hspace{-0.3in}&&  
  \nonumber
   \\  \hspace{-0.3in}&&  
 4\tau \left[x_1x_2+x-(1/4)x^2\right]_{\tau } I (\tau  )=(1+2\tau ) I (\tau )-1=
   \nonumber 
   {1- (1+2\tau )b_1(\tau )\over 2\tau } \simeq   
    \\  \hspace{-0.3in}&& \label{Eq:[2(x_1x_2+x)-(1/2)x^2]_tau  I A} 
     \\  \hspace{-0.3in}&&
        \tau (1-\tau +2\tau^2/3+\mathcal {O}(\tau ^3)) ~~(for ~\tau ~<1) \nonumber  ~~~;~~
   {1\over 2\tau }-{1 \over 12 \tau ^2}~~~~~~~~~(for ~\tau ~>1)
        \end{eqnarray} 
   This combination is used later in  the perturbative expansion terms. 
   \section{Strategy for Expansions \label{SEC:IV}}
   Our method is  illustrated by the inelastic case assuming at first for simplicity that the transmission coefficients in the leads  $T_{a,b,c}$ are small. In the case of the GS result, the dependence on the transmission coefficients in Eq. (\ref{Eq:EGSprod}) is then   governed by the factor   $\prod_e(1+2\tau T_e)^{-1/2} $.  A related factor  $\prod _e (1-\tau xT_e)(1+2\tau x_1T_e)^{-1/2}(1+2\tau x_2T_e)^{-1/2}$ appears in the VWZ model  as the product  contribution  to the integrand of Eq. (\ref{Eq:time}), which  approximately gives such a factor for small and moderate $ T_e$.  This suggests that one should avoid power expansions based on $2\tau T_e<1$ as a small parameter, since such expansions mostly have a narrow radius of convergence for the GS case. A far better expansion is obtained by the following procedure.\\
   Since  $x_1+x_2=1-x$ and with the notation $T'_e=T_e/(1+2\tau T_e)$ one can write %
\footnote{ The following expansion assumes that $2 x_1 x_2 \tau T_e<1$, typically.  Equation (\ref{Eq:[2(x_1x_2+x)-(1/2)x^2]_tau  I A}) suggests  the characteristic value of $x_1x_2$ is of order $\mathcal{O} (4\tau )^{-1})$ or less in typical integrals for $\tau >1$.  One can verify that  deviations lead to small and negligible contributions to the correction terms even in the limit of very strong transmission $T_e\simeq 1$.  }
    after   breaking out the factor $(1+2\tau T_e)^{-1/2}$ :       
            \begin{eqnarray}
   \hspace{-0.3in}&&  
    {(1-\tau T_ex) \over (1+2\tau x _1T_e)^{1/2}(1+2\tau x_2T_e)^{1/2}} 
\equiv  {(1-\tau T_ex) \over (1-2\tau T_e^{\prime }x +4 x_1x_2\tau ^2T_eT_e^{\prime })^{1/2}} \times  (1+2\tau T_e)^{-1/2}
           \label {Eq:elasticantennaexpansionAppA}
      \end{eqnarray}
   The first term of this product  is then exponentiated. The individual contributions in the exponent are expanded  assuming sufficiently small expansion parameters
         \begin{eqnarray}
   \hspace{-0.3in}&&  
      \prod _e   {(1-\tau T_ex) \over (1+2\tau x _1T_e)^{1/2}(1+2\tau x_2T_e)^{1/2}} = \label {Eq:basicexpansion}  
       \\   \hspace{-0.3in}&&  
\exp  \left\{\sum _e \left[  \ln (1-\tau xT_e)-(1/2)\ln \left(1-2\tau T_e^{\prime }x +4 x_1x_2\tau ^2T_eT_e^{\prime }\right)\right]\right\}\times   \prod_e  (1+2\tau T_e)^{-1/2}   \simeq   \nonumber 
      \\   \hspace{-0.3in}&&  
    \exp  \{-2\tau ^2  \sum_e T_eT_e' \left(x_1x_2 +x-(1/4)x^2\right)\} \times   \prod_e (1+2\tau T_e)^{-1/2}  \nonumber
         \end{eqnarray}
         In this expansion the linear terms in $x_1x_2$ and $x$ regroup naturally with the term in $x^2$ for small $\tau $. For $\tau >1$ this is no longer the case, but terms of second  order  and higher in these expansion quantities are negligible in the relevant regions of integration as compared to the linear terms. \\
          Note that the expansion is unrelated to the issue of whether the resonances are overlapping.\\
           Corresponding approximations can be made for the lead
            correction factors $g_{\tilde C_{1,2}}(\tau )$ which according to Eq. (\ref{Eq:C1,2-1tilde}) are
         \begin{eqnarray}
 \hspace{-0.3in}&&
 g_{\tilde C^{1}}((x_1,x_2,x; \tau ) =  \label{Eq:gC1taua}
 \left({x _1\over 1+2\tau T_a x_1}+{ x _2\over 1+2\tau T_a x _2}+{x \over 1-\tau T_a x }\right)       \times  \left({x _1\over 1+2\tau T_c x_1}+{ x _2\over 1+2\tau T_cx _2}+{x \over 1-\tau T_cx }\right)
   \end{eqnarray}
   \\
       \begin{eqnarray}
 \hspace{-0.3in}&&
 g_{\tilde C^{2}}((x_1,x_2,x;\tau ) =  \label{Eq:gC2taua}
 \Big[  {x _1(1+2\tau x_1)\over (1+2\tau T_a x _1)(1+2\tau T_b x _1)}+{\ x _2(1+2\tau x _2)\over (1+2\tau T_a x _2)(1+2\tau T_b x _2)}+{x (1-\tau x )\over (1-\tau T_ax) (1-\tau T_bx) }  \Big] 
   \end{eqnarray}
  Assuming as previously that no single decay channel dominates contribution from  the lead channels and extracting an  overall factor $(1+2\tau T_{a,b,c})^{-1}$ one finds a qualitatively different contribution, which now contains   terms which are linear in $\tau T_{a,b,c}$ which is of significance for the behavior of the correlation functions  for small $\tau $. This is contrary to the quadratic dependence on $\tau ^2T_e^2$ in the sum $\tau ^2\sum _eT_e^2$ of Eq. (\ref{Eq:basicexpansion}).  The next order terms in the transmission coefficients  of the leads are readily included, but do not qualitatively change the  results.  In accordance with the previous assumption that no channel dominates and for simplicity only the linear contribution from the leads is retained here.
 The correction term to this order  in the integrand  becomes   for the lead factor  $  g_{\tilde C^{1}}(\tau )$. 
   \begin{eqnarray}
 \hspace{-0.3in}&&
   2\tau   g_{\tilde C^{1}}((x_1,x_2,x;\tau ) \label {Eq:gC1}\simeq  2\tau (1+2\tau  T_a)^{-1}(1+2\tau  T_c)^{-1}   [ 1+ 4\tau(T_a+T_c) 
   (x_1x_2+x-x^2/4)+...]
 \end{eqnarray}

        For $ g_{\tilde C_{2}}(\tau )$  the leading terms in $\tau T_{a,b}$  are  with $\tau '=\tau /(1+2\tau )$
      \begin{eqnarray}
 \hspace{-0.3in}&&
  g_{\tilde C^{2}}(x_1,x_2,x;\tau )   \simeq  (1+2\tau )  (1+2\tau  T_a)^{-1}(1+2\tau  T_b)^{-1}  \label {Eq:gC2}
   [ 1-4\tau '\Big(1-
(1  +\tau ) (T_a+T_b) \Big) 
      (x_1x_2 +x-x^2/4)+...]  
   \end{eqnarray}
 The factors given in  Eqs. (\ref{Eq:gC1taua}-\ref{Eq:gC2taua}) with Eqs. (\ref{Eq:gC1}) and (\ref{Eq:gC2}), respectively, will be used in the following integrands.\\
   
   The relevant terms weighted in the VWZ integral therefore occur as  exponents proportional to $x_1x_2 +x-x^2/4$  characteristically  scaled by a multiplying term dependent on the transmission coefficients. When  the linear approximation  to the lead correction is valid over the important regions of the integrals we can exponentiate them and  combine the results with those of Eq. (\ref {Eq:basicexpansion}).  Introduced into the integrand of Eq. (\ref{Eq:identitiesA}) gives 
    \begin{eqnarray}
 \hspace{-0.3in}&&  
 \tilde C^1_{aa;cc}(\tau  ) \simeq  \left<S_{aa}\right> \left<S_{cc}^*\right>   2\tau  T_aT_c(1+2\tau  T_a)^{-1}(1+2\tau  T_c)^{-1}   \prod_e (1+2\tau T_e)^{-1/2} \label{Eq:C1,2-1tildeapprox1}
 \Big[ \exp \{-\kappa _1 \left(x_1x_2 +x-(1/4)x^2\right)\}\Big]_{\tau}  I(\tau )~~~~
    \\ \hspace{-0.3in}&& 
with ~~ \kappa _1= -4\tau(T_a+T_c) 
+2\tau ^2  \sum_e T_eT_e'  \label{Eq:kappa-1} 
 \end{eqnarray}
 and
   \begin{eqnarray}
 \hspace{-0.3in}&&  
  \tilde C^2_{ab;ab} (\tau  )\simeq   (1+2\tau ) T_aT_b  (1+2\tau  T_a)^{-1}(1+2\tau  T_b)^{-1}  \prod_e (1+2\tau T_e)^{-1/2} \Big[ \exp \{ -\kappa _2\left(x_1x_2 +x-(1/4)x^2\right)\}\Big]_{\tau } I(\tau )  ~~~~~~~ \label{Eq:C1,2-1tildeapprox2}
     \\ \hspace{-0.3in}&& 
with ~~  \kappa _2 =4\tau '-4\tau ' 
(1  +\tau ) (T_a+T_b)  +2\tau ^2  \sum_e T_eT_e'    \label{Eq:kappa-2} 
  \end{eqnarray}
This approximation is expressed here in a global form valid for all values of the time variable $\tau $.  It will now be evaluated in detail in different regions.\\
\section{\label{SEC:V} Explicit Results for the Approximation  }
         \subsection{Small and moderate $\tau$ with $\mathcal{O}(\kappa _{1,2})<max [1,\tau ]$}
         From Eqs. (\ref{Eq:C1,2-1tilde}) and (\ref{Eq:[2(x_1x_2+x)-(1/2)x^2]_tau  I A}) this gives expanding   Eq. (\ref{Eq:C1,2-1tildeapprox1}) for the   exponent smaller than unity \\
            \begin{eqnarray}
   \hspace{-0.3in}&&  \label{Eq:smalltauweakcoupling}
    \tilde C^1_{aa;cc} (\tau )  \simeq   \left<S_{aa}\right> \left<S_{cc}^*\right>2\tau T_aT_c  (1+2\tau  T_a)^{-1}(1+2\tau  T_c)^{-1}   \prod_e (1+2\tau T_e)^{-1/2} \times 
      \\ \hspace{-0.3in}&& 
\left[{1-b_1(\tau )\over 2\tau } + \left(\tau  (T_a+T_c)  \nonumber 
 -{1\over 2}\tau ^2   \sum_e T_eT_e' \right){ \left(  1-(1+2\tau ) b_1(\tau ) \right) \over 2\tau ^2 }\right] 
           \end{eqnarray}
          The restrictions from the integration region are automatically takien into account.\\
The corresponding result for $  \tilde C^2_{ab;ab} (\tau  ) $ from Eq. (\ref{Eq:C1,2-1tildeapprox2}) is 
    \begin{eqnarray}
   \hspace{-0.3in}&&  \label{Eq:C-2(tau)smalltau}
 \tilde C^2_{ab;ab} (\tau ) \simeq  T_aT_b (1+2\tau  T_a)^{-1}(1+2\tau  T_b)^{-1} \prod_e (1+2\tau T_e)^{-1/2} \times      
      \\ \hspace{-0.3in}&& \nonumber
        \left[1+\left(\tau (1+\tau )(T_a+T_b)-{1\over 2}\tau ^2(1+2\tau )   \sum_eT_eT_e'\right) {\Big(   1-(1+2 \tau )  b_1(\tau ) \Big) \over 2\tau ^2}\right]
            \end{eqnarray}
       For 
      $\tau>1 $ the Dyson function in Eq. (\ref {Eq:b-1}) is very small with $b_1(\tau )\simeq 1/ 12)\tau ^{-2}$ and gives a negligible contribution. Within the expansion region         $ \tilde C^1_{aa;cc} (\tau >1 ) $ becomes
                   \begin{eqnarray}
   \hspace{-0.3in}&&  \label{Eq:C1smalltauweakcoupling}
    \tilde C^1_{aa;cc} (\tau >1 )  \simeq 
     \\ \hspace{-0.3in}&& \nonumber 
      \left<S_{aa}\right> \left<S_{cc}^*\right> T_aT_c (1+2\tau  T_a)^{-1}(1+2\tau  T_c)^{-1}   \prod_e (1+2\tau T_e)^{-1/2} \times 
\left(1 +  \Big(T_a+T_c 
 -{1\over 2}\tau \sum_e T_eT_e' \Big)(1+\mathcal{O}(\tau ^{-1}))\right)
   \\ \hspace{-0.3in}&& \nonumber 
   \rightarrow  
    \left<S_{aa}\right> \left<S_{cc}^*\right> T_aT_c (1+2\tau  T_a)^{-1}(1+2\tau  T_c)^{-1}   \prod_e (1+2\tau T_e)^{-1/2} \times 
\left(1 + T_a+T_c 
 -{1\over 4} \sum_e T_e\right)~ for ~\tau \rightarrow \infty ~and~\sum_e T_e<2
           \end{eqnarray}
      
The corresponding result for $  \tilde C^2_{ab;ab} (\tau  ) $  is 
    \begin{eqnarray}
   \hspace{-0.3in}&&  \label{Eq:C-2(tau)smalltauA}
 \tilde C^2_{ab;ab} (\tau >1) \simeq  
 T_aT_b(1+2\tau  T_a)^{-1}(1+2\tau  T_b)^{-1} \prod_e (1+2\tau T_e)^{-1/2} \times 
         \left(1+{1\over 2}\Big(T_a+T_b-\tau    \sum_eT_eT_e' \Big)(1+\mathcal{O}(\tau ^{-1})) \right)
         \\ \hspace{-0.3in}&& \nonumber 
   \rightarrow    T_aT_b (1+2\tau  T_a)^{-1}(1+2\tau  T_b)^{-1}   \prod_e (1+2\tau T_e)^{-1/2} \times 
\left(1 +{1\over 2}( T_a+T_b )
 -{1\over 4} \sum_e T_e\right)~ for ~\tau \rightarrow \infty ~and~\sum_e T_e<2
            \end{eqnarray} 
           Note that the  limit $\tau \rightarrow \infty $ applies only  to the case of non-overlapping levels.
            \end{widetext}
 \subsection{Moderate and large $\tau $ with $\mathcal{O}(\kappa _{1,2})>max(1,\tau )$}
In this limit the previous expansion breaks down and the problem is dominated by the exponent appearing in Eqs. (\ref{Eq:C1,2-1tildeapprox1}) and (\ref{Eq:C1,2-1tildeapprox2}). 
In its simplest form the lead terms are weak.  Then the unobserved channels determine the correlations.  
 The typical integrand has the factor $\prod_e (1+2\tau T_e)^{-1/2}\left[  \exp ~[-\kappa _{1,2} \left(x_1x_2 +x-(1/4)x^2\right)]\right] $. 
It contains in addition the contribution 
  to the integrand  from  the time-factor  which  formally corresponds to an additional channel with $T_e=1$ which can be expanded as for any $T_e$ with $\tau'=\tau/(1+2\tau )$.
 These two decreasing factors set the scale for  contributions such that  for the weakly coupled inelastic case these come effectively from the region for which
\begin{eqnarray}
 \hspace{-0.3in}&&  \label{Eq:taucondition}
 \left(x_1x_2 +x-(1/4)x^2\right) \sim \mathcal {O}\left({1\over \tau \tau ^{\prime}+\tau ^{2}\sum_eT_eT_e^{\prime }}\right)
  \end{eqnarray} 
  Since $x_1+x_2+x=1$ and $x_{1,2}$ enter symmetrically we can take $x_1>x_2$ and multiply by 2 . When the integration variable in Eq. (\ref{Eq:taucondition}) is limited to small values, we have $x_1\rightarrow 1$,  while  $x_2$ and $x$ are both effectively small with  $x_2<<1; x<<1$. In this limit the integration can be carried out explicitly. Note however that for $\tau >1$ we have $x<1/\tau $, while the integration region for $x_2$ is limited only by $x_2<1/2$. The full  integration region corresponds approximately to a long narrow rectangle for large $\tau $.  Equation (\ref{Eq:taucondition}) can be used to chose $\tau $  such that the contributions come  effectively  only from a region for which both variables are much smaller than $1/\tau $. These conditions should be introduced into the basic integration as given in Eq. (\ref{Eq:identitiesA}) but in addition the explicit time-dependence of the integrand should be expressed  correspondingly and exponentiated.
  \newline
  Under these conditions the integral over $x_1$ is immediate. The integrand over $x_2,x$ concentrates well within the integration region which can be extended to $\infty $ for both $x_2,x$  without restrictions.  From Eqs. (\ref{Eq:C1,2-1tildeapprox1}) and (\ref{Eq:C1,2-1tildeapprox2})  this corresponds to the explicit evaluation of the terms $\Big[ \exp  -\kappa _{1,2}\left(x_1x_2 +x-(1/4)x^2\right)\Big]_{\tau } I(\tau ) $ using that the time-factor is  $~(1-\tau  x ) ~ (1+2\tau x _1)^{-1/2}(1+2\tau x _2)^{-1/2}$ $\simeq  $   
  $(1+2\tau)^{-1/2}$$\exp [-2\tau \tau '  \left (x_1x_2 +x-(1/4)x^2\right)]$  to a good approximation in this region. Consequently,  as described in Appendix  \ref{App:Integral} we have in terms of the integral $I(a,\tau )$ of Eqs. (\ref{Eq:I(a,tau)}) and (\ref{Eq:largetauintegral})
 \begin{widetext}
      \begin{eqnarray}
 \hspace{-0.3in}&&  
 \tilde C^1_{aa;cc}(\tau  ) \simeq     \left<S_{aa}\right> \left<S_{cc}^*\right>2\tau  T_aT_c (1+2\tau  T_a)^{-1}(1+2\tau  T_c)^{-1}   \prod_e (1+2\tau T_e)^{-1/2} \label{Eq:C1,2-1tildeapprox12}
 \times {1\over 2}    \left({\pi \over 2\tau ^2  +(1+2\tau )( \kappa _{1}-1/2)} \right)^{1/2}  
   \\ \hspace{-0.3in}&&  \nonumber
{\rm where }~~ \kappa _1= -4\tau(T_a+T_c) 
+2\tau ^2  \sum_e T_eT_e'  \label{Eq:kappa-1A} \end{eqnarray}
The corresponding expression for $  \tilde C^2$ is  closely similar :
   \begin{eqnarray}
 \hspace{-0.3in}&&  
  \tilde C^2_{ab;ab} (\tau  )\simeq  T_aT_b (1+2\tau ) T_aT_b (1+2\tau  T_a)^{-1}(1+2\tau  T_b)^{-1}  \prod_e (1+2\tau T_e)^{-1/2} 
  \label{Eq:C1,2-1tildeapprox22}
 \times {1\over 2}  \left({\pi \over 2\tau ^2  +(1+2\tau )( \kappa _{2}-1/2)} \right)^{1/2}  
   \\ \hspace{-0.3in}&&  \nonumber
 ~{\rm where }~
  \kappa _2 =4\tau '-4\tau ' 
(1  +\tau ) (T_a+T_b)  +2\tau ^2  \sum_e T_eT_e'    \label{Eq:kappa-2A} 
  \end{eqnarray}
  In the region  $\tau >1$ these expressions simplify neglecting contributions of order $\tau ^{-1}$ in the correction terms.
   \begin{eqnarray}
 \hspace{-0.3in}&&  
 \tilde C^1_{aa;cc}(\tau  ) \simeq  
  \left<S_{aa}\right> \left<S_{cc}^*\right> T_aT_c (1+2\tau  T_a)^{-1}(1+2\tau  T_c)^{-1}   \prod_e (1+2\tau T_e)^{-1/2} \label{Eq:C1,2-1tildeapprox12taularge}
  \times \left({\pi \over 2}\right)^{1/2}  \left({1\over 1-4(T_a+T_c)  +2\tau   \sum_e T_eT_e' }\right)^{1/2}  ~~~~
\end{eqnarray}
     \begin{eqnarray}
 \hspace{-0.3in}&&  
  \tilde C^2_{ab;ab} (\tau >>1  )\rightarrow    T_aT_b (1+2\tau  T_a)^{-1}(1+2\tau  T_b)^{-1}  \prod_e (1+2\tau T_e)^{-1/2} 
  \label{Eq:C1,2-1tildeapprox22largetau}
 \times \left({\pi \over 2}\right)^{1/2}  \left({1\over 1-2(T_a+T_b)  +2\tau   \sum_e T_eT_e' }\right)^{1/2}  ~~~~~~~~
   \\ \hspace{-0.3in}&&  \nonumber
 { \rm For~ asymptotically~large~ \tau ~these ~expressions ~ become}
      \\ \hspace{-0.3in}&&  
 \tilde C^1_{aa;cc}(\tau  \rightarrow \infty  ))\rightarrow  
  \left<S_{aa}\right> \left<S_{cc}^*\right>  T_aT_c (1+2\tau  T_a)^{-1}(1+2\tau  T_c)^{-1}   \prod_e (1+2\tau T_e)^{-1/2} \label{Eq:C1-1tildeapprox1tauinfty}
  \times \left({\pi \over 2}\right)^{1/2}  \left({1\over 1-4(T_a+T_c)  +   \sum_e T_e }\right)^{1/2}  ~~~~~
     \\ \hspace{-0.3in}&&  
      \tilde C^2_{ab;ab} (\tau \rightarrow \infty  )\rightarrow    T_aT_b (1+2\tau  T_a)^{-1}(1+2\tau  T_b)^{-1}  \prod_e (1+2\tau T_e)^{-1/2} 
 \times \left({\pi \over 2}\right)^{1/2}  \left({1\over 1-2(T_a+T_b)  +  \sum_e T_e }\right)^{1/2}  ~~~  \label{Eq:C2-1tildeapprox2tauinfty}
  \end{eqnarray}
    \subsection{An Interpolation Formula}
   The behavior of the correlation functions is most easily visualized in terms of  a closed analytical expression valid for the entire parameter space, even at the cost of a slightly reduced accuracy.  The previous expressions cover most regions of the time variable, but the small and large $\tau $ expansions do not quite overlap. The interpolation between these two regions is easily obtained  on interpolating by hand. These expressions depend on the   combination   $-2(1-\tau ' )(T_a+T_b)+\tau   \sum_eT_eT_e' $   for $ \tilde C^2_{ab;ab} (\tau  )$ and on a similar one,  $ -2(T_a+T_c) 
+\tau  \sum_e T_eT_e'  $, in the case of $ \tilde C^1_{aa;bb} (\tau  )$. Except for a minor effect at small $\tau $ produced by the linear dependence on the lead parameters $T_{a,b,c}$, the correction term as compared to the GS expression is a smooth, monotonously  decreasing positive function of  $\tau \sum_eT_eT_e'$.  For  $ \tilde C^2_{ab;ab} (\tau  )$, a  simple interpolated expression  is obtained by replacing the small $\tau $ expansion in Eq. (\ref{Eq:C-2(tau)smalltau}) which contains  a term   of the type $1-y$  with a small parameter $y$ by an expression of the type $(1+2y)^{-1/2}\simeq 1-y.$
\\
This gives, over the entire region, the interpolated value
  \begin{eqnarray}
   \hspace{-0.3in}&&  \label{Eq:C-2(tau)int}
     \tilde C^2_{ab;ab} (\tau ) \simeq T_aT_b  (1+2\tau  T_a)^{-1}(1+2\tau  T_b)^{-1} \prod_e (1+2\tau T_e)^{-1/2} \times \nonumber
     \\ \hspace{-0.3in}&& 
     \left\{1+2 \Big(-\tau (1+\tau )(T_a+T_b)+{1\over 2}  \tau ^2 (1+2\tau )   \sum_eT_eT_e'\Big) {\Big(   1-(1+2 \tau )  b_1(\tau ) \Big) \over 2\tau ^2 }\right\}^{-1/2}  \label{Eq:C2limit}
      \end{eqnarray}
In this form  the same expression can be used both for isolated and for overlapping levels.             This expression describes the global shape of the VWZ result very well over the entire range, including its absolute value even at large  $\tau $ for which the correlation functions become extremely small. It differs there from the more exact expressions             by an overall factor $2/\sqrt{\pi }=1.128..$ 
           for  $\tau \sum_eT_eT_e'>>1$. 
            For very   large $\tau $ the VWZ factor  in Eq. (\ref{Eq:C2limit}) is dominated by the sum over those open channels for which $\tau T_e>>1$ since $ 2\tau   \sum_eT_eT_e' \rightarrow  \sum_eT_e $.
            
  In the region  $\tau >1$ nearly the same expression is valid also for the elastic correlation function $  \tilde C^1_{aa;cc} (\tau )$  both for isolated and for overlapping levels. The dependence of the correlation functions $ \tilde {C}^{1,2}$  on the level correlation function $b_1(\tau )$ is nearly negligible in this region and is indicated in Eqs. (\ref{Eq:C2limit}) and (\ref{Eq:C-1(tau)int}) only to emphasize the continuity of the expressions.  
   For the region $\tau <1$, however, the expression (\ref{Eq:smalltauweakcoupling}) should be used, however, so as to correctly incorporate the effects of the level correlations. One has
               \begin{eqnarray}
   \hspace{-0.3in}&&  \label{Eq:C-1(tau)int}
     \tilde {C}^1_{aa;cc} (\tau ) \simeq   \left<S_{aa}\right> \left<S_{cc}^*\right> T_aT_c (1+2\tau  T_a)^{-1}(1+2\tau  T_c)^{-1} \prod_e (1+2\tau T_e)^{-1/2} \times \nonumber
     \\ \hspace{-0.3in}&& 
     \left\{1+2 \Big(-2\tau (1+\tau )(T_a+T_c)+{1\over 2}  \tau ^2 (1+2\tau )   \sum_eT_eT_e'\Big) {\Big(   1-(1+2 \tau )  b_1(\tau ) \Big) \over 2\tau ^2 }\right\}^{-1/2}  \label{Eq:C1limit} ~~{\rm for~\tau ~>1}
      \end{eqnarray}
   Here the level correlation function $b_1(\tau )$ is nearly negligible for $\tau >1$, but it is included above to emphasize the generality of the expressions  (\ref{Eq:C2limit}) and (\ref{Eq:C-1(tau)int}) for all values of $\tau $.  
 The implications of these results will be discussed in the next section.
      \end{widetext}
\section{Discussion \label{SEC:Disc}}
The previous results demonstrate that  the chaotic amplitude correlations of the VWZ model can be accurately described by a closed analytical  expression which~is at a the same time both simple and transparent. \\  To avoid misunderstandings, we remind the reader that the results have been obtained assuming for simplicity that no single channel dominates the sum of the partial widths nor the sum of their squares.  This assumption is not basic and can be  generalized.  In  particular, the case of a small number of channels  can be obtained using the same technique even without this assumption.  The price is a considerably more complicated discussion. 
 \\
  The analytical representation reveals  that  the gross shape of the correlation function  is dominated by a characteristic product factor which is identical to the one in Eq. (\ref{Eq:EGSprod})  for the amplitude correlations in  the GS solution of the EGS model.  This property stands out  particularly well in the case of the inelastic correlation function $ \tilde C^2_{ab;ab} (\tau )$.
For the EGS model it is a direct consequence of the  assumption that the partial width amplitudes are random, i. e., have individually Gaussian distributions  uncorrelated between channels.  The result for the VWZ model  demonstrates that in this case these amplitudes have become dynamically random to a high degree. This strong dependence on the width amplitudes and their distributions means that the correlation functions are particularly sensitive to any deviations from chaotic conditions in  this sector and this will be so for the VWZ model as well.  This is of considerable practical interest, since it opens the possibility of direct investigations of   the sensitivity of systems approaching full chaoticity to amplitude distributions which are not normal.  The consequences of possible modifications have been investigated in the case of the GS solution \cite{DIE12}. Such results  are immediately relevant the VWZ case as well.\\
  
Corresponding observations are also valid for  the elastic correlation function $ \tilde C^1_{aa;cc}(\tau )$ which for  $\tau >1$  rapidly converges to the same expression as for $ \tilde C^2_{ab;ab} (\tau )$ but for an overall factor.  The elastic and inelastic correlation functions  differ for $\tau <1$ mainly due to the dominance of the level correlation function $b_1(\tau )$ in Eq. (\ref{Eq:b-1}) in $ \tilde C^1_{aa;cc}(\tau )$ in this region; a natural form for this function  is the Dyson function which explicitly is a consequence of  the VWZ model \cite{VER86}, but introduced phenomenologically "{\it ad hoc}' in the EGS model and its solutions. From the present approximate  VWZ results one observes directly that  its global properties are the important ones: the level repulsion on the scale of the level spacing and  the detailed  normalization of the "correlation hole" in the spacing distribution which it produces. 
These level spacing correlations rapidly become unimportant for larger $\tau $.  In this region the correlation function explores predominantly  the average structure of individual broadened levels.   \\
   A general feature of the  VWZ elastic correlation function is the  linear dependence on $\tau $ near the origin. It is presently clear from Eq. (\ref {Eq:C1smalltauweakcoupling}) that this is a consequence of  the level repulsion by which a level creates an adjacent "hole" corresponding to exactly one missing level.  
    Since the Fourier transform at $\tau =0$ of the correlation function $\tilde {C}_{aa;cc}^1 (\tau  ) $ is the energy integral over the entire range, the joint contribution of the level and its "hole"  vanishes in this case. This effect  has previously been well investigated in the context of the EGS model\cite{GOR02};  the present result reveals a more detailed, yet still simple, picture  of the interplay of the level distribution and the transmission coefficients.
\\

  It is frequently taken for granted in the literature that the correlation functions  decrease exponentially for large $\tau $  for strongly overlapping levels.   The present analytical VWZ description shows clearly that this is only approximately correct as is also the case for  the GS expression.   The correct asymptotic dependence of the correlations in the VWZ  model is  an inverse power law $\tau ^{-(M/2+2)}$ with  $M$ is the number of open channels $e$ in agreement with the  behavior  previously suggested by the  EGS model\cite{GOR02}.  With increasing $\tau $  a power $\tau ^{-1/2} $ is added successively every time $\tau T_e$ becomes larger than $1$ for any channel.  For experimental  tests of the properties of such systems,  it is then useful to consider not only the the main global exponential decay parameter $\sum _eT_e$, but also $\sum _eT_e^2$ which governs the onset of the deviation from the exponential law. 
 \\
   
            The differences of the  2-point function for the exact VWZ solution to that of the EGS model appear mainly as 3 characteristic effects. 
\\
    
1.  The 2-point function of the VWZ model differs  from the one of the simplified EGS model by an overall modulation function given in Eqs. (\ref{Eq:C-2(tau)int},\ref{Eq:C-1(tau)int}). It is a considerable experimental challenge to display this function explicitly owing to the high statistical precision required. Its effects are  very interesting from a theoretical viewpoint, however.  They can be investigated in detail both using exact numerical VWZ results, using our approximations as well as using numerical  computer simulations of chaotic systems. This modification stands out particularly clearly in case of the inelastic correlation function $ \tilde C^2_{ab;ab} (\tau )$.  It has a  typical behavior  apparent from the expressions in Eqs. (\ref{Eq:smalltauweakcoupling},\ref{Eq:C-2(tau)smalltau}) and (\ref{Eq:C1,2-1tildeapprox12},\ref{Eq:C1,2-1tildeapprox22}). \\
For $\tau >1$ the dominant term is modified by a factor $(1+2\tau \sum _eT_eT_e')^{-1/2}$, which decreases monotonically with $\tau $. In the asymptotic power law limit of very large $\tau $  this factor becomes approximately   $( 1+\sum _eT_e)^{-1/2}$ both for isolated and for overlapping levels
\footnote{A frequent practical situation is that of open channels which divide effectively into two groups, one with transmission coefficients contributing both to $\sum _eT_e$ and $\sum T_e^2$ at the relevant  $\tau $ and another one with small transmission coefficients contributing only to $\sum _eT_e $, 
but with a negligible  effective contribution to $\sum _eT_e^2$.  These latter weak channels give a contribution indistinguishable from the one for true absorption with an exponential contribution to the correlation function with a decay parameter $\Lambda _{abs} =\sum _eT_e^{small }$.  In this case these weak channels do not contribute to 
terms $\sum _eT_eT_e'$ in Eqs. (\ref{Eq:C1,2-1tildeapprox12taularge},\ref{Eq:C1,2-1tildeapprox22largetau}) nor 
to the asymptotic expression containing $\sum _eT_e$, nor do they effectively occur in the asymptotic power expansion. This observation in no way invalidates the asymptotic power law. It only means that the region of exploration should be sufficiently large with  $\tau T_e>1$ for \underline {all}   of the channels. }. 
 This asymptotic limit   is of little practical importance since the characteristic constant factor is overshadowed by the power law variation $\tau ^{-(M/2+2)}$.  \\
 
 2.  The VWZ model has a typical suppression factor $<S_{aa}><S_{cc}>=(1-T_a)^{1/2}(1-T_c)^{1/2}$ in the case of the correlation function $C^1$ contributing to elastic correlations as  apparent already in the general expression (\ref{Grassman}) as well as in the detailed expressions (\ref{Eq:smalltauweakcoupling},\ref{Eq:C1,2-1tildeapprox12}) and (\ref{Eq:C-2(tau)int},\ref{Eq:C-1(tau)int}). This factor has the consequence that the correlation function for 2 elastic amplitudes vanishes in the extreme  limit when one or the other of the transmission coefficients reaches the limit of very strong transmission, i.e., for $T_{a,c}=1$.
   Similarly, in this same limit the elastic autocorrelation function is partly suppressed, but only by about  30\%.  The GS solution used here for comparison has the factor  $1$ instead of the factor  $<S_{aa}><S_{cc}>$, but the limit $T_{a,c}=1$ is outside its range of its applicability as discussed in Ref. \cite{DIE12}.  The difference  originates in  the unitarity constraint, which is not imposed on the EGS model.
   For smaller $T_{a,c}$ the effect of this difference becomes minor.  The GS expression can be improved by introducing this VWZ factor phenomenologically.   One should however realize that  the case  of very strong transmission is  of limited practical interest.
    This multiplicative factor in the elastic channels is the single most important difference between the VWZ and EGS models  for chaotic amplitude correlations. \\
 
3. For $\tau <1$ the modulation function depends explicitly on the level spacing distribution also in the inelastic case. In the region of very small $\tau $ this inelastic factor is  $[1+\tau (T_a+T_{b})-1/2 \tau ^2\sum _eT_e^2+..]$.  At small $\tau $ this leads  to a minor initial increase  of the modulation factor with a maximum  returning to $1$ near $\tau =2(T_a+T_{b})/\sum _eT_e^2$  and thereafter joining the previously described monotonic decrease for $\tau >1$.
 A  closely related factor modifies  the correlation function $  \tilde C^1_{aa;cc}(\tau ) $, but it is then difficult to disentangle  from the level spacing correlation function.  \\
 
The present results provide an encouraging structure for the approximate description of higher order correlation functions for chaotic and nearly chaotic systems for which  exact results are known only in special cases.    The close correspondence the present results from the VWZ model to those of the  EGS model suggests that the latter should provide a reliable, albeit approximate,  guide to this situation. 
 This feature also suggests more generally that the simplified  EGS model is a valid laboratory for exploring the sensitivity of various features which govern the approach to the fully chaotic situation of the VWZ model. In particular, the physics of the sensitivity of the correlation functions  to symmetry violations can be reliably clarified using the simpler EGS model.

 \begin{widetext}
\section {Acknowledgements}
I am indebted to Dr. B. Dietz and Prof. A. Richter for raising questions that have led to the present investigation. In particular, the  numerical results by Dr. Dietz on the VWZ model originally drew  my attention to the regularities of the VWZ model deviations from those of the  GS expression.   I am also indebted to Prof. A.W. Weidenm\"uller for useful  comments.

\appendix
 \section{The   2-point Correlations in the EGS Model: Time Representation \label{App:EGS} }
The EGS model gives the following results for the present case of normally distributed  partial width amplitudes  in each of the uncorrelated channels \cite{GOR02,DIE12}, i.e., partial widths with Porter-Thomas distributions \cite{POR56}:       
 \begin{eqnarray}
   \hspace{-0.3in}&&    
   \tilde{C}[S_{a b}S^*_{a b}](\tau)= T_aT_b \Pi _{e;ab}(    \tau)
~for~ a\neq b~~~~~~~~~~~~~~~~~~~~~ (inelastic~ autocorrelations) \label {Eq:EGSprod}  \\
  \hspace{-0.3in}&&   \nonumber 
 \tilde{C}[S_{a a}S^*_{cc}](\tau)=  T_aT_c \Big( \Pi _{e;ac}( \tau)  -\rho _0   b(\tau) \Pi _{e;a}(\tau/2)\Pi _{e;c}(\tau/2)\Big) ~~(elastic ~correlations)  \label{Eq:2pointfulltime} 
  \end{eqnarray}
for $\tau>0$ and $0$ for $\tau <0$.   

    Here
\begin{eqnarray}
   \hspace{-0.3in}&& 
   \Pi _{e;\alpha ,\beta, ..}(\tau )= (1+2\tau T_{\alpha })^{-1}(1+2\tau T_{\beta })^{-1}..\prod_e(1+2\tau T_e)^{-1/2} 
    \end{eqnarray}
   for $\alpha \neq \beta $. It  is multiplied by $3$ for $\alpha =\beta $ in the case a normal distribution. \\
     The exponential approximation to the time-variation holds if the variance of the total width obeys the inequality:
 $(2\pi)^2\tau^2<(\Gamma -<\Gamma >)^2>=\tau ^2 \sum _eT_e^2<1$ . 
 \section{The integral $I(a,\tau )$ in the large $a$ limit\label{App:Integral}}
   Consider an integral of the type $I(a,\tau )$, symmetrical in $x_1$ and $x_2$, for large values of the parameter $a>>max (1,\tau ) $ to leading order neglecting corrections of order $1/a$.
    \begin{eqnarray}
 \hspace{-0.3in}&&  
I(a,\tau )\equiv \label{Eq:I(a,tau)}
 \\ \hspace{-0.3in}&&  \nonumber
 \int ^{\infty }_0 dx _1\int ^{\infty }_0 dx _2 \int ^{\, min (1,1/\tau ) }_0 dx \delta (x_1+x_2+x-1) 
  {x |x _1-x _2| \over  
(x _1x _2)^{1/2}( x +2x _1)^2(x +2x _2)^2}
 \exp \left[-a [x_1x_2+x-(1/4)x^2]\right] ~~~~~~~~
  \end{eqnarray}
For the case $x_1=1-x_2-x>x_2\simeq 1$ it follows that  the relevant region of integration over $x_2+x<<1$  can then be freely extended to $\infty $, since this gives negligible extra contributions. In this case the term $( x +2x _1)^2\simeq 4[1-2x_2-x]=4(x_1-x_2)$ which is the numerator term such that the ratio to a good approximations is $1$ which simplifies the discussion (an omitted term $(x_2+x/2)^2$ is assumed small compared to $1$). Denoting $x_2=y=u^2$ and $x=z=v^2$ with $\rho ^2=u^2+v^2$ and $u=\rho \cos \phi ; v=\rho \sin \phi $ gives on exponentiating the contributions associated with $x_1$,
   \begin{eqnarray}
 \hspace{-0.3in}&&  
I(a,\tau ) 
\simeq  {2\over 4}\int ^{\infty }_0 dy \int ^{\infty }_0 dz  {z \over  
y^{1/2}(z +2y)^2}
 \exp \left[-(a-1/2) [y+z]\right] =\nonumber
 \\ \hspace{-0.3in}&&  
 2 \int _0^{\infty } d\rho \int _0^{\pi /2} d\phi {\sin ^3\phi \over (1+\cos ^2 \phi )^2 }\exp [-(a-1/2)\rho ^2]
  \rightarrow  \left({\pi \over 2(2a-1) }\right)^{1/2}~~~~  \label{Eq:largetauintegral} {\rm for~a>>max (1,\tau ) }
 \end{eqnarray}
 where the the last step is obtained using
 $ \int _0^{\infty}dt\exp(-k t^2)=\left({\pi /4k }\right)^{1/2}~
 ;~\int ^{\pi /2 }_0 d\phi   \sin ^3 \phi  (1+\cos ^2\phi)^{-2}=1/2$
  \end{widetext}
    
    \end{document}